
\magnification=\magstep1
\baselineskip=24pt

\def\eqdef{\buildrel {\rm def} \over =}

\vskip 20pt
\centerline{\bf Conductance Fluctuations of Disordered Mesoscopic
Devices}
\centerline{\bf with Many Weakly-Coupled Probes}
\vskip 10pt
\centerline{{\it Martin R. Zirnbauer} \footnote* {On leave from: Institut
f\"ur Theoretische Physik, Universit\"at zu K\"oln, Germany}}
\vskip 10pt
\centerline{Wissenschaftskolleg zu Berlin, Germany}
\vskip 30pt

\centerline{ABSTRACT}
\bigskip

The conductance coefficients of disordered mesoscopic devices with $n$
probes are investigated within the noninteracting electron
approximation at zero temperature. The probes are eliminated from the
theoretical description at the expense of introducing nonlocal
boundary conditions at the position of the contacts. Conductors with a
large number of weak contacts are analyzed in detail. The
ensemble-averaged conductance coefficients are in this case given by the
so-called Hauser-Feshbach formula $\langle g_{ab} \rangle \sim p_a p_b$
where $p_a$ is the probability for emission of an electron into lead
$a$. The conductance energy autocorrelation function is shown to deviate
significantly from the conventional Lorentzian form and to have a negative
tail.\vskip 3cm
\noindent published in: Nucl. Phys. {\bf A560} (1993) 95-116
\vfill\eject
\baselineskip=13.4pt

\noindent{\bf 1. Introduction}\medskip

At temperatures in the sub-Kelvin range, disordered metallic
devices with a linear size $L$ of about $1\mu$m or less display
random but reproducible conductance fluctuations as a function
of an externally applied magnetic field $B$ [1].
It is now well understood [2] that these fluctuations,
whose amplitude is universally of the order of $e^2/h$,
originate from the phase-coherent and diffusive quantum motion
of electrons through the interior of the device,
the ``mesoscopic conductor''. The sample-specific
fluctuations with varying magnetic field are related, by an
ergodicity argument [3], to statistical fluctuations across
an ensemble of conductors characterized by
the diffusion constant $D$. To estimate the correlation
field $B_c$, one uses a semiclassical path-sum argument
and equates $L^2/D$ with the time it takes in order for
a typical diffusive electron path to acquire from the magnetic
field $B_{c}$ an extra quantum mechanical phase equal
to $\pi$. \medskip

The authors of Ref. [4] first pointed out that the value of
$B_{c}$ is sensitive to the contact apertures joining the
mesoscopic conductor to the current-voltage probes. They
noticed that the mean time an electron dwells inside the
conductor before escaping into the leads, is enhanced
over $\tau_{\rm diff} = L^2/D$ in the case of small apertures. Such an
increase in dwelling time amounts to an increase in the area enclosed
by a typical Feynman path, thereby leading to a reduction of $B_c$.
\medskip

Ref.~[4], while arriving at conclusions that are quite sound within
the specific model considered, fell short of realizing the need for
explicit introduction of a separate and {\it independent} time
scale in addition to the diffusion time $\tau_{\rm diff}$. This step
was made in Ref.~[5], where the importance of the {\it decay time}
$\tau_{\rm decay}$ for the mesoscopic conductor problem was recognized.
A lucid discussion of the physical meaning of $\tau_{\rm decay}$
is given in Ref.~[6]. The new time scale is set by the decay width
$\Gamma = \hbar / \tau_{\rm decay}$ for electron emission into one of
the leads. Ref.~[6] quotes the formula
	$$
	\Gamma = {\Delta \over 2\pi} \alpha N ,
	\eqno(1)
	$$
where $\Delta$ is the level spacing of the mesoscopic conductor,
$N$ is the number of open scattering channels at the Fermi
energy $E_F$ in the leads, and $\alpha$ is a dimensionless parameter
measuring the quality of the contacts. ($\alpha \simeq 1$ for a good
contact, and $\alpha \ll 1$ for a weak contact.) When $\Gamma \gg
E_c = hD/L^2$ or, equivalently, $\tau_{\rm decay} \ll \tau_{\rm
diff}$, it is the Thouless energy $E_c$ that defines the correlation
energy of the problem and $\Gamma$ becomes an irrelevant parameter.
On the other hand, in the opposite limit $\Gamma \ll E_c$, roles
are interchanged and it is $\Gamma$, not $E_c$, that sets the
relevant energy scale. Mesoscopic conductors satisfying the
inequality $\Gamma \ll E_c$ form the subject of the second half of
this paper and will be called ``decay-width dominated'' for short.
\medskip

Formula (1) shows that there exist three independent ways of
reaching the decay-width dominated regime: (i) reduce $N$ by
making the leads thinner; (ii) reduce $\alpha$ by degrading
the quality of the contacts; and (iii) if the conductor is
three-dimensional (resp. quasi-one-dimensional), reduce
$\Delta \sim L^{-d}$ relative to $E_c \sim L^{-2}$ by increasing
(resp. decreasing) $L$. The second option is pushed to the extreme
by placing {\it tunnelling barriers} at the contacts, that is
to say, potential barriers which electrons must tunnel through in
order to reach the conductor from one of the leads or vice versa.
\medskip

The purpose of the present paper is twofold. First, we wish to
formulate the mesoscopic conductor problem in a way which, like
the model of Ref.~[5], takes proper account of the existence of the
decay width $\Gamma$, but differs from Ref.~[5] in several respects:
it does not exploit the connection with scattering theory but utilizes
a more elementary expression for the conductance coefficients in terms
of Green's functions; it uses the standard
continuum Gaussian white noise model rather than Wegner's $n$-orbital
model; and it eliminates the leads by a procedure known as
R-matrix theory [7-10]. This procedure is particularly well suited
for conductors with tunnelling barriers, where a treatment in terms of the
scattering matrix seems awkward and may even be unfeasible. \medskip

The second purpose is to communicate some analytical results for the
conductance fluctuations of devices with tunnelling barriers and, more
generally, of decay-width dominated conductors in the ``locally-weak
absorption limit''. By this term we mean the limit in which the
electron emission rate per unit area of a contact becomes small and
the total area of all contacts grows large, while their product is
kept fixed. A special case of this limit are devices with a large
number of weak contacts. To our knowledge, the limit of locally weak
absorption has not been considered before in mesoscopic conductor
physics (although its nuclear physics analog has received some
attention [11]), perhaps because its experimental realizability is
not clear. Nevertheless, we find it instructive to consider this limit
since it (i) adds to our theoretical understanding of the phenomenon of
universal conductance fluctuations and (ii) stands out by some distinctive
features concerning correlations. In fact, we will show that the
functional form of the conductance energy autocorrelation function
deviates markedly from the conventional Lorentzian one in this case.
\medskip

An outline of the contents of this paper is as follows. Sect.~2 defines
the microscopic model and quotes the basic expression for the
conductance coefficients $g_{ab}$ $(a,b=1,...,n)$ in terms of advanced
and retarded Green's functions at the Fermi energy. In
Sect.~3, the problem of calculating $g_{ab}$ is reformulated. The
probes are eliminated and replaced by nonlocal boundary conditions on
the surfaces separating the mesoscopic conductor from the leads. In
Sect.~4, it is argued that for conductors with tunnelling barriers a local
approximation for the integral kernel defining the boundary conditions
may be used. This leads to the formulation of a phenomenological
model, whose ensemble-averaged conductance coefficients $\langle
g_{ab} \rangle$ are calculated in Sect.~5 and shown to be given by the
so-called Hauser-Feshbach formula known from the statistical theory of
nuclear reactions. Sect.~6 is concerned with the conductance
covariance function $\langle \delta g_{ab}(E_1) \delta g_{cd}(E_2)
\rangle$. This function turns negative as $|E_1-E_2|$ increases beyond
$\Gamma$ and satisfies the sum rule of having vanishing integral over
$E_2$ (or $E_1$). Our results are summarized in Sect.~7.
\medskip

I owe my understanding of many of the physical concepts and mathematical
tools underlying this article to Hans A. Weidenm\"uller. It is
therefore appropriate that the article be
dedicated to him on the occasion of his $60^{\rm th}$ birthday and
published in the present volume.
\bigskip

\noindent
{\bf 2. The model and its conductance coefficients}
\medskip

In this paper, three-dimensional disordered metallic devices of the
generic type will be considered. Such devices consist of a
mesoscopic conductor, roughly of
linear size $L$, and of $n$ leads joined to the
conductor by the contacts. Both the conductor and its leads
may have arbitrary geometrical shapes.
\medskip

Following common practice in mesoscopic physics, we assume that the
statistical features of the conductance coefficients of such a device,
at zero temperature, can be modelled by an ensemble of single-electron (or
mean-field) Hamiltonians of the general form
	$$
	H = {1 \over 2m} \bigl( {\bf p}-e{\bf A} \bigr)^2 + U +
	{\bf U}_m \cdot {\bf \sigma} + \bigl( {\bf p}-e{\bf A} \bigr)
	\cdot \bigl( {\bf U}_{SO} \times{\bf \sigma} \bigr) .
	\eqno(2)
	$$
The functions $U$, ${\bf U}_m$ and ${\bf U}_{SO}$ are taken to be
Gaussian white-noise potentials, with their strengths determined
by the mean-free times for potential scattering, $\tau$, magnetic
spin-flip scattering, $\tau_{m}$, and spin-orbit scattering,
$\tau_{SO}$, respectively [12]. The times $\tau_m$, $\tau_{SO}$
and the inverse cyclotron frequency $(e/m \times {\rm rot}{\bf A})^{-1}$ are
supposed to be long compared to $\tau$. To avoid complications that
would otherwise appear in Sects.~5 and 6, we assume the contacts to be
separated from each other by a distance of at least a few times the
elastic mean-free path $\ell = v_F \tau$. \medskip

The objects of the present investigation are the conductance
coefficients $G_{ab} = e^2/h \times g_{ab}$ ($a,b=1,...,n$),
which determine the current response
of the $n$-probe conductor to the applied electrostatic potentials.
Our starting point is a formula for $g_{ab}$ which originates from
linear response theory and can be found in Ref.~[13];
see also Ref.~[14]. To write it down, the following definitions
are needed. We take $G^+$ (resp. $G^-$) to be the retarded (resp.
advanced) one-particle Green's function of $H$ at the
Fermi energy $E_F$; formally:
	$$
	G^\pm(x,y) = \lim_{\varepsilon\to 0+} ~ \bigl(E_F \pm
	i\varepsilon - H \bigr)^{-1} (x,y) .
	$$
The limit $\varepsilon\to 0+$ exists because the ``openness'' of the
system makes the spectrum of $H$ absolutely continuous.
To keep the notation simple, we adopt the
convention that the symbols $x$, $y$ etc. comprise both position
coordinates and spin projection.
We introduce the conductivity tensor
	$$\eqalignno{
	\sigma_{\alpha\beta}(x,y) = ~
	&(v_\alpha G^+)(x,y)~(v_\beta G^-)(y,x) +
	(G^+ v_\beta)(x,y)~(G^- v_\alpha)(y,x) 			\cr
	+ &(v_\alpha G^+ v_\beta)(x,y)~G^-(y,x)
	+ G^+(x,y)~(v_\beta G^- v_\alpha)(y,x)	&(3{\rm a})	\cr}
	$$
where $v_\alpha$ is the $\alpha$-component $(\alpha=1,2,3)$
of the velocity operator
	$$
	{\bf v} = {\partial H\over\partial {\bf p}} = {1\over m}
	({\bf p}-e{\bf A}) + {\bf U}_{SO} \times {\bf\sigma} .
	\eqno(4)
	$$
For $a \in \{1, ..., n\}$, let $c_a$ be any cross section of lead $a$.
To account for spin, we introduce two identical
copies of $c_a$, one for each spin projection:
$c_a(\uparrow)$ and $c_a(\downarrow)$, and we set
$C_a = c_a(\uparrow) \cup c_a(\downarrow)$.
Eq.~(75) of Ref.~[13] can then be cast in the form $(a \not= b)$
	$$
	g_{ab} = - {\hbar^2 \over 4} \int_{C_a} d^2x \int_{C_b}
	d^2y ~ \sigma_{ab}(x,y) .
	\eqno(3{\rm b})
	$$
Here $\int_{C_a} d^2x$ stands for integration over $c_a$
and summation over spin, and, with $n_a$ the vector normal to $c_a$,
	$$
	\sigma_{ab}(x,y) = \sum_{\alpha,\beta}
	n_a^\alpha(x) n_b^\beta(y) \sigma_{\alpha\beta}(x,y) .
	\eqno(3{\rm c})
	$$
(Note that Ref.~[13] ignores the spin degrees of freedom
but the extension is straightforward.) \bigskip

\noindent{\bf 3. Elimination of the leads}\medskip

According to a standard assumption made in mesoscopic physics, the
properties of the conductance coefficients $g_{ab}$ are determined
primarily
by the phase-coherent diffusive motion of electrons inside the conductor
and to a lesser extent by the details of what happens in the leads. In
fact, one often takes the leads to be ``ideal'' or ``clean'' (although
in reality the contacts and the leads are usually made from the very
same dirty metal as is the mesoscopic conductor), i.e. one
approximates the motion in the leads by free motion. Such an
approximation is well justified if there is a clear geometrical division
between the conductor and the leads, as effected for example by small
contact apertures. In a situation where details of the motion in the
leads do not matter, it makes sense to try and eliminate the leads from
the theoretical formulation altogether. This can be done by a
procedure known as R-matrix theory [9] in nuclear physics. (Note,
however, that the roles played by the interior and the exterior of
configuration space will be interchanged relative to the nuclear
case.) It is applied in the present context as follows. \medskip

Let $a$ and $b$ two fixed elements of the set $\{1,...,n\}$.
Evaluation of the conductance coefficient $g_{ab}$ from Eq.~(3)
requires as input the Green's functions $G^+(x,y)$ and $G^-(y,x)$, and
their derivatives, for $x\in C_a$ and $y\in C_b$.
We will show how to construct $G^+$; $G^-$ can
then be obtained from the relation $G^-(y,x) = \overline{G^+(x,y)}$.
\medskip

We observe that the total configuration space, $V$, is partitioned by the
surfaces $C_c$ into the conductor space, $V_0$, and the lead spaces,
$V_c$ $(c = 1, ...,n)$. Suppose now that we are to calculate
$G^+(x,y)$ for $x\in C_a$ and $y\in V_0$. (We take $y$ to $C_b$ in
the end.) To begin, recall that $G^+(\cdot,y) \equiv \psi(\cdot)$
is a solution of $(E_F-H)\psi = 0$ on the region $V-\{y\}$ with
outgoing-wave boundary conditions. As as a first step, we take
$G_a^-(z,x)$ to be the advanced Green's function of $H$ (again
evaluated at $E_F$) with support contained in $V_a$ and subject
to the boundary condition $G_a^-(z,x) = 0$ for all $z\in C_a$ and
$x\in V_a$. We write $G_a^-(\cdot,x) \equiv G_x^-(\cdot)$
for short. If ${\cal D}$ denotes the ``covariant derivative''
	$$
	{\cal D} = \nabla - {ie\over\hbar}{\bf A}
	+ {im\over\hbar} {\bf U}_{SO}\times\sigma ,
	$$
current conservation implies the validity of the relation
	$$
	{\rm div} \left( \overline{G_x^-} \cdot {\cal D\psi}
	- \overline{{\cal D}G_x^-} \cdot \psi \right) = 0
	\eqno(5)
	$$
on all of $V_a$ with the exception of the point $x\in V_a$. The dots in
Eq.~(5) indicate summation over spin. Next, we integrate Eq.~(5) over the
region $V_a - \{x\}$ and use Green's theorem to obtain
	$$
	\psi(x) = - {\hbar^2\over 2m} \int_{C_a}
	\overline{({\cal D}_a G_x^-)(z)} \psi(z) d^2z,
	\eqno(6)
	$$
where $\hbar{\cal D}_a /i = m v_a$ with $v_a = \sum_\alpha n_a^\alpha
v_\alpha$ and $n_a$ is the vector normal to $c_a$ as before.
Now we apply $v_a$ to both sides of (6) and take the point $x$ to
$C_a$. Using the relation $\overline{G_x^-(z)} = G_a^+(x,z)$ we then get
	$$
	(v_a \psi)(x) = \int_{C_a} B_a(x,z) \psi(z) d^2z
	\quad (x\in C_a)
	\eqno(7)
	$$
where
	$$
	B_a(x,z) = {i\hbar\over 2}
	\Bigl( v_a G_a^+ v_a \Bigr)(x,z) \qquad (x,z \in C_a).
	\eqno(8)
	$$
Eq.~(7) relates the normal component of the covariant derivative of
$\psi$ on $C_a$ to the values and the derivatives
of $\psi$ on $C_a$. This relation
can be regarded as a boundary condition satisfied by $\psi$. We have
thus arrived at an {\it exact reformulation} of the original problem:
instead of solving the Schr\"odinger equation for $\psi$ on the total
space $V-\{y\}$ (with outgoing-wave boundary conditions), we may solve
$(E_F-H)\psi = 0$ on the restricted space $V_0-\{y\}$ supplemented with the
boundary condition (7). The latter procedure produces all the
information needed, viz. the values of $\psi \equiv G^+(\cdot,y)$
on $C_a$. \medskip

Returning finally to complete notation and the specific task posed by
Eqs.~(3a-c), we observe that the Green's functions
$G^\pm(p,q)$ for $p, q \in V_0$ are
computed by solving $\bigl( (E_F-H) G^\pm \bigr)(p,q) = \delta(p-q)$
together with the boundary conditions
	$$\eqalignno{
	(v_c G^+)(u,q) &=
	\int_{C_c} B_c(u,z) G^+(z,q) d^2z =
	\overline{(G^- v_c)(q,u)}		&(9{\rm a})	\cr
	(v_c G^-)(u,q) &=
	- \int_{C_c} \overline{B_c(z,u)} G^-(z,q) d^2z
	= \overline{(G^+ v_c)(q,u)}		&(9{\rm b})	\cr}
	$$
for all $u \in C_c$ $(c=1,...,n)$. At the end of the computation we
take $p \to x\in C_a$ and $q \to y \in C_b$.
Let us now decompose the operators $B_c$ as $B_c = {\rm Re}B_c + i {\rm
Im}B_c$ where both $({\rm Re}B_c)(x,y)$ and $({\rm Im}B_c)(x,y)$ satisfy
the hermitecity relation $O(x,y) = \overline{O(y,x)}$.
Eqs.~(9) then permit us to recast (3b) in the form
	$$
	g_{ab} = \hbar^2 \int_{C_a} d^2x \int_{C_b}
	d^2y ~ \bigl( ({\rm Re}B_a) G^+ \bigr) (x,y)
	\bigl( ({\rm Re}B_b) G^- \bigr)(y,x)
	\eqno(10{\rm a})
	$$
where ($c=a$ or $c=b$)
	$$
	\bigl( ({\rm Re}B_c) G^\pm \bigr) (x,y) =
	\int_{C_c} ({\rm Re}B_c)(x,z) G^\pm(z,y) d^2z .
	\eqno(10{\rm b})
	$$
\medskip

Of course, the usefulness of the above reformulation depends
critically upon the possibility of finding {\it adequate and
managable} approximations to the {\it nonlocal} integral
kernels $B_a(x,y)$ $(a=1,...,n)$. This possibility in turn is
contingent upon our ability to controll the Green's functions
$G_a^+$, see Eq.~(8). \medskip

In devising approximations to $B_a$, one should note that
${\rm Im}B_a$ corresponds, in a rough manner of speaking,
to reflection at $C_a$. It
is of minor importance and can often be neglected. On the other hand,
${\rm Re}B_a$ causes ``absorption'' (taking the point of view of
restricted configuration space $V_0$), that is to say, the loss of
probability due to emission of electrons into lead $a$. This latter
effect is crucial for determining the relevant time scales of the
mesoscopic conductor, and it must therefore be modelled properly.
\medskip

By its definition in terms of $G_a^+$ through Eq.~(8), $B_a$
depends on the location of the surface $C_a$, and the difficulty of
devising a good approximation varies with location. It is therefore
very fortunate that current conservation leaves complete freedom in
choosing the surfaces $C_a$, thereby permitting us to optimize the
choice. How to make an optimal choice is rather obvious if there
exists a clear geometrical division between the conductor and the
leads, and if the leads can be taken to be clean: one will then place
$C_a$ right at the contact of lead $a$. In this case, an approximation
which captures the essential features of $B_a$ is
	$$
	B_a(x,y) = \pi \sum_c
	W_a^c(x) \overline{W_a^c(y)} ,
	\eqno(11)
	$$
where the sum runs over all open scattering channels at $E_F$ in lead
$a$, and the amplitudes $W_a^c(x)$ are chosen phenomenologically to
fit the average rate of emission into the channels. This corresponds
to the approximation used in Ref.~[5] at the level of the S-matrix.
Another case where $B_a$ can be controlled with ease is analyzed in
the sequel.
\vfill\eject

\baselineskip=12.8pt
\noindent
{\bf 4. Reduced model for devices with tunnelling barriers}
\medskip

{}From now on we will restrict ourselves to the case of $n$-probe
conductors with {\it weak} contacts. In other words, emission into the
leads will be supposed to be inhibited by obstacles of some kind. For
definiteness, we imagine these obstacles in the form of tunnelling
barriers and add a barrier potential $U_B$ to the Hamiltonian $H$ of
Eq.~(2). Let us denote by $C_a'$ (resp. $C_a''$) the surface
separating the classically forbidden region under the barrier at
contact $a$ from the allowed region inside the conductor (resp. in lead
$a$) . An optimal choice (in the sense of the previous section) is
then to put $C_a = C_a'$ $(a=1,...,n)$. We adopt this choice. \medskip

The simplification that occurs for devices of such kind is that one
may use a {\it local} approximation to the kernels $B_a$. This is
intuitively clear and can be quantified by the following semiclassical
argument. \medskip

Let $a \in \{1,...,n\}$.
To construct $B_a$, we must first calculate $G_a^+(x,y)$ for
two points $x$ and $y$ under the barrier and close to $C_a$, then apply
the derivatives $v_a$ (see Eq.~(8)) and finally take $x$ and $y$ to
$C_a$. For simplicity, we will assume that the motion under the
barrier is determined mostly by the interplay of $U_B$ with the
kinetic energy $p^2/2m$ and neglect the influence of all
other terms in $H$. \medskip

By applying the stationary-phase approximation [15] to Feynman's path
integral for the propagator [16], we write $G_a^+(x,y)$ as a
sum over classical paths with energy $E_F$ connecting $y$ with $x$. The
contribution largest in magnitude comes from the path of shortest
length, which, for $U_B$ constant in the barrier region, would be
a straight-line
trajectory. This path, while generating the correct Green's function
singularity as the distance $|x-y|$ goes to zero, is uninteresting
here since it is an imaginary-time path without conjugate points and
its contribution to $G_a^+$ (resp. $B_a$) is purely real (resp. purely
imaginary). \medskip

An imaginary contribution to $G_a^+$
arises from the path that takes off from $y$ in
the direction of lead $a$, {\it bounces} off the surface $C_a''$ and
then heads straight for $x$. If $l_B$ denotes the thickness of the
tunnelling barrier and $\kappa_F$ the average value of
$\sqrt{2m(U_B-E_F)}/\hbar$ along this path, the path's action is
roughly given by
	$$
	S \sim 2i\hbar\kappa_F \sqrt{l_B^2 + |x-y|^2/4} .
	$$
There exist three more classical paths of a similar kind. These are
the paths that are reflected at the surface $C_a$ before heading for $C_a''$
and/or before arriving at $x$. They carry opposite sign factors, and
their combined contribution to the semiclassical approximation for
$G_a^+(x,y)$ vanishes as $x$ and/or $y$ approaches $C_a$, as it
should. \medskip

In summary, by the above argument
	$$
	{\rm Im} G_a^+(x,y) \sim - c_0
	d_a(x) d_a(y) e^{-2\kappa_F \sqrt{l_B^2 + |x-y|^2/4}}
	$$
where $d_a(\cdot)$ is distance from $C_a$, and $c_0$ is a positive
constant. Taking derivatives and
sending $x$ and $y$ to $C_a$, we obtain ${\rm Re}B_a(x,x) \sim
\exp(-2\kappa_F l_B)$, and we arrive at the important conclusion that
the range of ${\rm Re}B_a(x,y)$ is of the order of
$\sqrt{l_B/\kappa_F}$. This range is to be compared to the elastic
mean-free length, $\ell$, which is the smallest length scale of relevance
for the diffusive motion inside the conductor. Inserting reasonable
numbers ($\kappa_F^{-1} > l_B \sim 1 nm$ and $\ell > 10 nm$), we see
that $\ell \gg \sqrt{l_B/\kappa_F}$. It is therefore justified, for the
purpose of calculating the conductance coefficients, to
suppress the length scale $\sqrt{l_B/\kappa_F}$ and use a local
approximation for ${\rm Re}B_a(x,y)$. \medskip

We are thus led to the following {\it reduced model}.
We take reduced position space, $V_0$, to consist of the interior of the
device only. $\partial V_0$ denotes the total boundary of $V_0$,
and $C \eqdef \cup_{a=1}^n C_a$ (resp. $\partial V_0-C$) its
conducting (resp. insulating) part. We introduce a phenomenological
function $\beta: \partial V_0 \to {\bf R}$ characterizing
the penetrability of the barrier; $\beta(x) \not= 0$ for
$x \in C$, and $\beta(x) = 0$ for $x \in \partial V_0-C$.
$\beta$ is assumed to be spin-independent (neglecting Zeeman splitting
in the presence of a magnetic field, and magnetic spin-flip and
spin-orbit scattering under the barrier). We impose the boundary
condition
\footnote* {Although Dirichlet's boundary condition might seem more
realistic for the insulating part of the surface, we feel that the
internal consistency of the reduced model is enhanced by using
boundary conditions of the form (12) everywhere. In any case, in
the diffusive regime we intend to study, the conductance coefficients
and their statistical properties are affected by the conductor-insulator
boundary conditions only in a minor way.}
	$$
	(v_n\psi)(x) = \beta(x)\psi(x)
	\quad ({\rm all}~~x\in \partial V_0).
	\eqno(12)
	$$
Here $v_n$ is the component of ${\bf v}$, Eq.~(4), normal (in the
spatial sense) to
$\partial V_0$. We take the Hamiltonian $H$ to be a linear operator
of the general form of Eq.~(2), restricted to act on functions $\psi$
supported on $V_0$ and subject to the boundary condition (12).
We define $G^+(x,y)$ to be the kernel of the operator
$(E_F-H)^{-1}$ with the boundary condition (12), and we set
$G^-(x,y) = \overline{G^+(y,x)}$. The conductance coefficients
are then calculated from
	$$
	g_{ab} = \hbar^2 \int_{C_a} d^2x \int_{C_b}
	d^2y ~ \beta(x) G^+(x,y) \beta(y) G^- (y,x).
	\eqno(13)
	$$
This completes the definition of the reduced model.
\medskip

Of course, the Hamiltonian $H$, which was a hermitean operator in
the original full space (including the leads, and with the usual
square-integrability condition), becomes {\it non-hermitean} upon
introduction of the boundary conditions (12), unless $\beta$ vanishes
identically. This non-hermitecity is an inevitable and, in fact,
essential feature of any reduced description of the present type, and
is particularly evident from the following alternative formulation
of the reduced model.
\medskip

Let $\delta_C$ be Dirac's $\delta$-distribution with uniform support on
$C$, {\rm i.e.}
	$$
	\int_{V_0} \delta_C f d^3x = \int_C f d^2x ,
	$$
and introduce an effective Hamiltonian
	$$
	H_{\rm eff} = H - {i\hbar\over 2} \beta\delta_C .
	\eqno(14)
	$$
Then $G^+$, as defined above, is also the Green's function of the
linear operator $H_{\rm eff}$ acting on functions $\psi$ that satisfy
the generalized Neumann boundary conditions
	$$
	v_n \psi \big |_{\partial V_0} = 0 .
	\eqno(15)
	$$
To verify this statement, integrate $(E-H_{\rm eff})\psi = 0$
along an infinitesimal piece of curve intersecting $C$ at a right
angle, and use the boundary condition (15). From the continuity of
$\psi$ across $C$, the boundary condition (12) is then recovered.
The extra term $-i\hbar\beta\delta_C/2$ in $H_{\rm eff}$ has an
interpretation as an ``absorptive contact potential'', giving rise to
the loss of probability which is caused by the escape of flux through
$C$. Note that the escape rate at $x\in C$, per unit area and spin
projection, is given by $\beta(x)|\psi(x)|^2$ for a state $\psi$ with
amplitude $\psi(x)$. \bigskip

\noindent
{\bf 5. Average conductance coefficients of decay-width dominated
devices in the locally-weak absorption limit} \medskip

Given an ensemble of one-particle Hamiltonians of the form specified
in Sect.~2, we will now calculate for the reduced model of Sect.~4
the ensemble average $\langle g_{ab} \rangle$ ($a\not= b$) for
decay-width dominated conductors in the limit of locally weak absorption.
\medskip

Recall first the definition of decay-width dominated conductors by
the inequality $\tau_{\rm diff} \ll \tau_{\rm decay}$ where
$\tau_{\rm diff} = L^2/D$ and, with ${\rm vol} = \int_{V_0} d^3x$,
	$$
	\tau_{\rm decay}^{-1} = {\rm vol}^{-1} \int_C \beta d^2x .
	$$
Recall also the definition of the locally-weak absorption limit by
$\beta\to 0$ and $\int_C d^2x \to\infty$ with $\int_C \beta d^2x$
kept fixed. A more quantitive condition for this limit to be attained
is
	$$
	\beta(x) \ell^2 \ll \int_C \beta d^2x \quad {\rm for}
	~~{\rm all} ~~ x\in C
	\eqno(16)
	$$
where $\ell = v_F\tau$ is the elastic mean-free path for potential
scattering. The inequality (16) is motivated by the observation
that, since $\ell$ is the lower cutoff length for diffusive motion,
$\int_C d^2x$ should be compared to $\ell^2$. \medskip

To compute $\langle g_{ab} \rangle$ from Eq.~(13), we require the
ensemble average
	 $$
	P(x,y) = \langle |G^+(x,y)|^2 \rangle =
	\langle G^+(x,y)G^-(y,x)\rangle
	$$
for $x \in C_a$ and $y \in C_b$. We will now calculate this quantity
in the prescribed limit. To begin, let $x$ and $y$ be any two points in $V_0$.
The inequality $\tau_{\rm diff} \ll \tau_{\rm decay}$ implies
that an electron moving diffusively inside the mesoscopic
conductor, traverses the distance $L$ many times before being
absorbed at one of the contacts (that is to say, before being
emitted into one of the leads). The density $|G^+(x,y)|^2$ of a
stationary state sustained by a source at $y$ therefore becomes
independent of $x$ upon ensemble averaging, with two exceptions.
The first occurs whenever $x$ is within a distance of $\ell$ or less from
one of the contacts, where wave functions, and consequently
$P(x,y)$ as well, are depressed by absorption in general. However,
the condition (16) for absorption to be locally weak precisely means that
this depression can be neglected under the present circumstances. The
second exception occurs when $x$ lies within a distance of $\ell$
or less from the source point $y$. In this case, the value of
$P(x,y)$ is modified by {\it coherent backscattering}, as is
well known from weak localization physics [17]. Summarizing this
paragraph, we write
	$$
	P(x,y) = P_d + P_c(x,y)
	\eqno(17)
	$$
where $P_d$ is the constant part of $P$ and $P_c$ is the short-ranged
backscattering correction. The sign of $P_c(x,x)$ depends on the
relative strengths of potential scattering, magnetic
spin-flip scattering, spin-orbit scattering and the magnetic field
[12], but its magnitude is always comparable to $P_d$. If the
arguments presented in support of Eq.~(17) seem too heuristic, a
technical derivation is given in Appendix A. \medskip

To proceed, recall that we assume the contacts $a$ and $b$ to be
separated from each other by a distance of at least a few times $\ell$.
By Eq.~(13) we then need to
know $P_d$ but not $P_c(x,y)$. Multiplying both sides of Eq.~(17)
with $\beta(x)$, integrating over $\int_C d^2x$, and dividing by
$\int_C \beta d^2x$, we have
	$$
	P_d = \int_C d^2x~ \beta(x) \bigl( P(x,y) -
	P_c(x,y) \bigr) / \int_C \beta d^2x .
	\eqno(18)
	$$
The contribution to the right-hand side of (18) from $P_c$ is
negligible under the condition (16), since
	$$
	\int_C d^2x ~ \beta(x) P_c(x,y) \simeq
	\ell^2 \beta(y) P_c(y,y) \ll P_d \int_C \beta d^2x .
	$$
The remaining term can be calculated by using the identity
	$$
	i\hbar \int_C d^2y ~ G^+(x,y)\beta(y)G^-(y,x)
	= \bigl( G^- - G^+ \bigr) (x,x),
	\eqno(19)
	$$
which is a consequence of the boundary condition (12) and
of current conservation. Taking the
ensemble average on both sides and using
	$$
	\langle \bigl( G^- - G^+ \bigr)(x,x) \rangle
	= 2\pi i\nu ,
	$$
where $\nu$ (independent of $x$) is the local density of states,
we obtain
	$$
	P_d = {2\pi\nu\over \hbar} \left( \int_C \beta d^2x \right)^{-1}.
	$$
We insert this relation into the ensemble-averaged version of
Eq.~(13). Upon introduction of the quantities
	$$
	\Delta^{-1} = \nu\times {\rm vol}, \qquad
	\Gamma = \sum_{a=1}^n \Gamma_a, \qquad
	\Gamma_a = {\hbar\over {\rm vol}} \int_{C_a} \beta d^2x
	\qquad (a = 1, ...,n),
	$$
where $\Delta^{-1}$ is the total density of states (counting spin),
$\Gamma$ the total decay width, and $\Gamma_a$ the partial
decay width for emission into lead $a$, the
expression for $\langle g_{ab} \rangle$ takes the form
	$$
	\langle g_{ab} \rangle = {2\pi\over\Delta}
	\times {\Gamma_a \Gamma_b \over \Gamma}.
	\eqno(20)
	$$
This is the analog of what is called the ``Hauser-Feshbach formula''
in the statistical theory of nuclear reactions [18]. Note the
interpretation of the ratio $\Gamma_a / \Gamma = \int_{C_a} \beta
d^2x / \int_C \beta d^2x$ as the probability for emission of an
electron from the mesoscopic conductor into lead $a$. Note also that
in comparison with the Thouless formula [19], the Thouless energy
has been replaced by $2\pi\Gamma_a\Gamma_b/\Gamma$. \medskip

Let us mention in passing that the validity of the Hauser-Feshbach formula
is not confined to the locally-weak absorption limit. As a matter of fact,
all that is needed in order for $\langle g_{ab} \rangle$ to have
the factorized form of Eq.~(20), is the {\it long
dwelling time} of a decay-width dominated conductor,
causing the processes of entry from lead $a$ (or $b$) and emission
into lead $b$ (or $a$) to be uncorrelated. Thus, formula (20)
remains valid for strong absorption if the partial decay widths
$\Gamma_a$ are replaced by more complicated, nonlinear expressions
in $\beta$. We will not elaborate upon this point here. \bigskip

\noindent
{\bf 6. Conductance covariance function of decay-width dominated
devices in the locally-weak absorption limit}\medskip

Considering devices of the same special kind as before, we will now
calculate the correlation function
	$$
	\langle g_{ab}(E_F+\delta E) g_{cd}(E_F) \rangle
	$$
for $a\not= b$ and $c\not= d$. This will be done for the reduced model
of Sect.~4, taking $\beta$ to be {\it energy-independent}.
\footnote* {The justification for neglecting variations of $g_{ab}$
due to changes in the penetrability coefficient $\beta$ - which are
rather strong in a microscopic model with tunnelling barriers - is
that such variations do not concern us here. What is
measured in experiments with variable magnetic fields are the
disorder-induced {\it statistical} variations of $g_{ab}$, and it is
therefore the latter that we wish to calculate.} Eventually, we will
specialize to the three universality classes [20,17,12] that are
known to exist for the
mesoscopic conductor problem. Recall [21] that these are denoted by I
(orthogonal class: potential scattering), IIa (unitary class: magnetic
field), IIb (unitary class: magnetic spin-flip scattering) and III
(symplectic class: spin-orbit scattering). \medskip

With all the preparations made in Sect.~5, we can now be rather brief.
Proceeding as before, we are led to consider the ensemble average
	$$
	\langle
	G^+(x_a,x_b;E_1) G^-(x_b,x_a;E_1)
	G^+(x_c,x_d;E_2) G^-(x_d,x_c;E_2)
	\rangle
	\eqno(21)
	$$
for $x_a \in C_a$, ..., $x_d \in C_d$. Here it is necessary to
distinguish cases. Let us first assume that the indices $a$, $b$, $c$
and $d$ are all {\it mutually different} and, as before, contacts are
separated from each other by a distance of at least a few times the elastic
mean-free path $\ell$. Then the argument of Sect.~5 goes through without
change and expression (21) can be approximated with negligible error by
a constant, independent of $x_a$, ..., $x_d$. Introducing the
conductance covariance function
	$$
	C_{ab,cd}(E_1,E_2) = \langle g_{ab}(E_1) g_{cd}(E_2) \rangle
	- \langle g_{ab}(E_1) \rangle \langle g_{cd}(E_2) \rangle
	$$
and using again Eq.~(19), we find
	$$
	C_{ab,cd}(E_1,E_2) = \langle g_{ab}(E_1) g_{cd}(E_2)\rangle
	R(E_1,E_2),
	\eqno(22{\rm a})
	$$
where
	$$
	R(E_1,E_2) = {\langle {\rm Im}G^+(E_1)\times {\rm Im}G^+(E_2)\rangle
	\over \langle {\rm Im}G^+(E_1)\rangle
	\langle {\rm Im}G^+(E_2) \rangle} - 1,
	\eqno(22{\rm b})
	$$
and $G^+(E) = \overline{G^-(E)}$ is the trace of the Green's function
of the Hamiltonian $H$ with the boundary condition (12). \medskip

If, on the other hand, at least two indices in (21) coincide, or if
contacts are separated by a distance less than $\ell$, a complication
arises. Consider for definiteness the case $a=d$. There now arise
contributions to the double integral $\int_{C_a} d^2x_a \int_{C_a}
d^2x_d $ from the region $|x_a-x_d| < \ell$. In this region, the value of
the otherwise constant expression (21) is modified by cross
contractions between $G^+(x_a,x_b;E_1)$ and $G^-(x_d,x_c;E_2)$. In
diagrammatic language, these give rise to impurity ladders of the same
``cooperon'' type that cause the backscattering correction discussed
earlier. (For $a = c$ diffuson ladders appear, too.) Because of the
condition $|x_a-x_d| < \ell$, the additional terms lead to contributions
carrying the factor $\ell^2 \int_{C_a} \beta^2 d^2x$, which is to be
compared to the regular contribution, Eq.~(22a), carrying the factor
$(\int_{C_a}\beta d^2x)^2$. Thus the terms missing in Eq.~(22a) are
small if
	$$
	\ell^2 \int_{C_a} \beta^2 d^2x \ll
	\left( \int_{C_a} \beta d^2x \right) ^2
	\quad (a=1, ..., n).
	\eqno(23)
	$$
Note that this condition is much more restrictive than (16). In
summary, in order for Eq.~(22) to be correct, we must either suppose
that the indices $a$, $b$, $c$ and $d$ all differ from one another
(and $|x_i-x_j| > \ell$ for all $x_i \in C_i$, $x_j \in C_j$ ($i\not=
j$)) or else impose the stronger condition (23). \medskip

Eq.~(22) reduces the problem of calculating $C_{ab,cd}(E_1,E_2)$ to a
problem in ``level statistics'' for an open system. Note that in
contrast to Ref.~[19], where a somewhat similar relation appears,
$C_{ab,cd}$ is here expressed {\it directly} in terms of $R(E_1,E_2)$ and
{\it not} as a double integral over energy. The possibility to do
so is offered by the limit under consideration. \medskip

The calculation of (22b)
is still not easy in general but in the locally-weak absorption limit
and for each of the universality classes, it reduces to a problem whose
solution is known. For the following it will be convenient to
represent $G^+(E) = \overline{G^-(E)}$ alternatively by $G^+(E) = {\rm
tr} (E-H_{\rm eff})^{-1}$ with $H_{\rm eff} = H -
i\hbar\beta\delta_C/2$
and the Neumann boundary condition (15). \medskip

Consider the matrix element $M_{kl}$ of $\hbar\beta\delta_C$
between two eigenstates $\psi_k$ and $\psi_l$ of $H$:
	$$
	M_{kl} = \hbar \int_C ~ \overline{\psi_k(x)}
	\beta(x) \psi_l(x) d^2x.
	$$
The phases of $\psi_k$ and $\psi_l$ are randomized over a correlation
length of the order of the elastic mean-free path $\ell$. We therefore
expect
	$$
	\langle M_{kl} \rangle = \delta_{kl} {\hbar\over {\rm vol}}
	\int_C \beta d^2x = \delta_{kl}~\Gamma
	$$
and, by the law of large numbers,
	$$
	{\rm var}M_{kl} \sim O \left( \int_C d^2x / \ell^2 \right)^{-1} .
	$$
These estimates mean that $\hbar\beta\delta_C$ acts {\it effectively}
as a multiple of the unit operator under the condition (16). In other
words, for the purpose of calculating $R(E_1,E_2)$ we are permitted to
approximate $G^+(E)$ in Eq.~(22b) by ${\rm tr}(E+i\Gamma/2-H)^{-1}$.
To avoid possible misconceptions, we stress that the approximation
$\hbar\beta\delta_C \simeq \Gamma \times 1$ must not be
overinterpreted. For the case $\Gamma \gg \Delta$, which is included
in the limit where our considerations apply, the operator
$\hbar\beta\delta_C$ does cause appreciable mixing between states
close in energy. However, such mixing {\it does not modify the
correlation properties} of the spectrum. In particular, it does not
reduce the spectral rigidity. The approximation is therefore
justified when used for the purpose of calculating correlation
functions such as $R(E_1,E_2)$. \medskip

For the special case $\Gamma = 0$, the two-level correlation function
resulting from the substitution $G^+(E) \to {\rm
tr}(E+i\Gamma/2-H)^{-1}$ in Eq.~(22b) has been calculated exactly
for each universality class by Efetov [22] in his work on the level
statistics of small metallic particles. It is straightforward to
extend Efetov's calculation to $\Gamma\not= 0$. The result for
$R(E_1,E_2)$, obtained by performing this analytic continuation, is
quoted in Appendix B. \medskip

Here we refrain from considering the general case but specialize to
$\Gamma \gg \Delta$, the case of an ``open'' system, which is also
accessible (without Efetov's results) via standard diagrammatic
perturbation theory. Using either the exact result of Appendix B,
or more simply the perturbative cooperon-diffuson expansion - see
Eq.~(33) of Ref.~[19] - one obtains
	$$
	R_k(E_1,E_2) = - {c_k \Delta^2 \over 4\pi^2} ~ {\rm Re} \left(
	E_1-E_2 +i\Gamma \right)^{-2} ,
	\eqno(24)
	$$
where the distinction between universality classes is contained in the
coefficients $c_{\rm I} = 16$, $c_{\rm IIa} = 8$, $c_{\rm III} = 4$
and $c_{\rm IIb} = 2$. Insertion of (24) into (22a) and use of Eq.~(20)
gives
	$$
	C_{ab,cd}(E_1,E_2) = c_k p_a p_b p_c p_d ~
	{ 1 - (E_1-E_2)^2/\Gamma^2 \over
	(1 + (E_1-E_2)^2/\Gamma^2 )^2 }
	\eqno(25)
	$$
where $p_a = \Gamma_a / \Gamma$, $p_b = \Gamma_b / \Gamma$ etc. \medskip

The universal numbers $c_k$ ($k = {\rm I}, ...,{\rm III}$) reflect the
influence of symmetries on the conductance fluctuations. Breaking of
time-reversal invariance (${\rm I}\to {\rm IIa}$ or ${\rm III}\to {\rm
IIb}$) causes a reduction by a factor of 2, and breaking of
spin-rotation invariance (${\rm I}\to{\rm III}$ or ${\rm IIa}\to{\rm
IIb}$) causes a reduction by a factor of 4. These reduction factors
are not new but were already observed in [19]. \medskip

What is more striking about Eq.~(25) is the non-Lorentzian dependence
of $C_{ab,cd}$ on $(E_1-E_2)/\Gamma$ and the validity of the sum rule
	$$
	\int_{\bf R} C_{ab,cd}(E,E') dE' = 0 ,
	\eqno(26)
	$$
which is seen most easily from Eq.~(24) and Cauchy's theorem. We
emphasize that the result (26) does not hold in general but is a
special feature of the locally-weak absorption limit. (It is true, however,
for any value of the decay width $\Gamma$ of decay-width dominated
conductors; see Appendix B.) To understand this feature, we recall that
the rigidity of the energy spectrum of an isolated system is preserved
for an open (but decay-width dominated) system satisfying the
condition (16): energy levels are, roughly speaking, shifted by a
common amount $E \to E - i\Gamma/2$. The negative tail, or
``correlation hole'', of $R(E_1,E_2)$ for $|E_1-E_2| > \Gamma$
is a direct consequence of this rigidity. \medskip

The conductance fluctuations result from Eq.~(25) by setting $E_1 =
E_2$:
	$$
	\langle \delta g_{ab} \delta g_{cd} \rangle
	= c_k p_a p_b p_c p_d .
	$$
These fluctuations are universal in the sense that they are sensitive -
leaving aside the dependence on symmetry - only to the {\it geometry} of
the $n$-probe conductor as expressed by the coefficients $p_a$
$(a=1,...,n)$. However, this is not a universality in the strictest
sense of the word since $p_a = \Gamma_a / \sum_{k=1}^n \Gamma_k$
does become smaller as the total number $n$ of contacts increases. Strict
universality is recovered by dividing the contacts into two groups,
say ``left'' (L) and ``right'' (R), and summing over $a, c \in L$
$(a\not= c)$ and $b, d \in R$ $(b\not= d)$.
\vfill\eject
\noindent{\bf 7. Summary}\medskip

This paper was in two parts. In the first part, we addressed the
practical question of how to calculate the conductance coefficients
$g_{ab}$ $(a,b=1,...,n)$ from the linear-response formula (3a-c).
Supposing that the statistical properties of $g_{ab}$ are determined
mostly by the diffusive motion of electrons inside the mesoscopic
conductor, it is natural to try and eliminate the leads from the
theoretical description in favour of suitable boundary conditions
imposed on the contact surfaces $C_a$ $(a=1,...,n)$. This idea is
usually implemented in a very schematic fashion. Refs.~[2-4] invoke the
condition $P_d = 0$ (resp. $\nabla_n P_d = 0$) on the conducting (resp.
insulating ) part of the surface, at the level of contructing the
diffuson propagator $P_d$. Such a condition, while reasonable for thick
leads and good contacts, is too crude for thin leads and/or weak
contacts. (The authors of Ref.~[4] are of course aware of this
limitation. This is why for the thin-lead case they adopt the trick of
making a somewhat arbitrary division of the leads into clean and dirty
regions.) \medskip

Inspection of Eqs.~(3a-c) shows that what we need to do is to
calculate the covariant derivatives $(v_a G^\pm)(x,y)$ for $x \in
C_a$ $(a=1,...,n)$. In Sect.~3 we restated the problem of making this
calculation on the total space (including the leads) as an {\it
exactly equivalent but reduced} problem formulated on the bounded
domain of the mesoscopic conductor. The reduced formulation involves a
set of integral kernels $B_a$, see Eq.~(8), which define boundary
conditions expressing the {\it covariant derivative} of $G^\pm$ normal
to $C_a$ in terms of the {\it values} of $G^\pm$ on $C_a$
$(a=1,...,n)$. These {\it nonlocal} boundary conditions cause the
reduced (or effective) Hamiltonian to be {\it non-hermitean} and {\it
energy-dependent}. $B_a$ is determined exclusively by the properties
of electronic motion in the leads and is therefore easy to controll
when the
leads are clean; see Eq.~(11). (We did note translate the boundary
conditions (9) with $B_a$ given by (11) into a corresponding boundary
condition for the diffuson propagator $P_d$. How this is done is
described at length for a different microscopic model in Ref.~[5];
see also Sect.~7 of Ref.~[23] for some useful mathematical
details.) \medskip

Another example where $B_a$ can be controlled are conductors with
tunnelling barriers. The semiclassical argument given in Sect.~4 shows
that a local approximation to $B_a$ may be used in this case. This
argument led to the formulation of a phenomenological model, whose
diffuson propagator $P_d$ satisfies
	$$
	- \hbar D~\bigl( \nabla^2 P_d \bigr) (x,y) = 2\pi\nu~\delta(x-y)
	$$
with the boundary condition $- D\nabla_n P_d = \beta P_d$ on the surface
$C$ of the conductor. \medskip

In the second part of the paper, the phenomenological model of Sect.~4
was analyzed for decay-width dominated conductors in the locally-weak
absorption limit. The technical simplification occurring for
decay-width dominated conductors is that the spectral expansion of
$(-\hbar D\nabla^2)^{-1}$ is dominated by the smallest eigenvalue,
$\lambda_0$, of $-\hbar D\nabla^2$ (with the specified boundary
condition) in this case. $\lambda_0$ and the next eigenvalue, $\lambda_1$, are
given by
	$$\eqalign{
	\lambda_0 &\simeq \Gamma =
	{\hbar\over {\rm vol}} \int_C \beta d^2x,	\cr
	\lambda_1 &\simeq \Gamma + {\rm const}\times E_c
	\qquad (E_c = hD/L^2).		\cr}
	$$
Note that for a three-dimensional mesoscopic conductor the limit
$\lambda_0 \ll \lambda_1$ is approached by keeping the conducting part
of the surface constant in size and making $L$ {\it larger}.
\medskip

The locally-weak absorption limit is defined technically by the
condition (16). It guarantees the correlation properties of the
eigenvalues of the effective Hamiltonian $H_{\rm eff} = H -
i\hbar\beta\delta_C/2$ to be the same as those of $H-i\Gamma/2$. In
other words, the eigenvalues of the Hamiltonian $H$ of the isolated
system are simply shifted, roughly speaking, by a constant amount
$-i\Gamma/2$. \medskip

By expressing the conductance fluctuations as a density-of-states
correlation function in Eq.~(22) - which is rigorously justified in
the prescribed limit and for $a$, $b$, $c$, $d$ all mutually different
- we obtained
	$$
	\langle \delta g_{ab}(E_1) \delta g_{cd}(E_2) \rangle \sim
	{ 1 - (E_1-E_2)^2/\Gamma^2 \over
	(1 + (E_1-E_2)^2/\Gamma^2 )^2 }.
	$$
The interesting feature of this expression is that it turns negative
at $|E_1-E_2| > \Gamma$. The negative tail can be understood as being
a consequence of the spectral rigidity of $H$. \medskip

Finally, let us translate the correlation energy $\Gamma$ into the
corresponding magnetic-field correlation scale $B_c$ measurable in
experiments on a single sample. To do that, we will follow a procedure
described in Ref.~[4]. A difference in magnetic field ${\bf B} \to
{\bf B} + \delta {\bf B}$
couples into the equation for the diffuson propagator by the
substitution $\nabla \to \nabla - ie\delta {\bf A}/\hbar$ where
$\delta {\bf B} = {\rm rot}\delta {\bf A}$. We need to calculate the
correction to $\lambda_0$ in lowest order of $\delta{\bf B}$.
Perturbation theory gives $\lambda_0 \to \lambda_0 + {\rm const}
\times E_c (\delta \phi/\phi_0)^2$ where $\delta\phi/\phi_0$ is the
change in magnetic flux through the conductor, measured in units of the
flux quantum $\phi_0 = h/e$. By equating the correction to $\Gamma$,
we obtain
	$$
	B_c L^2 \sim \phi_0 \sqrt{\Gamma/E_c} .
	$$
Thus, in comparison with the case of thick-lead conductors with good
contacts, $B_c$ is reduced by a factor of $\sqrt{\Gamma/E_c}$.
\bigskip

{\bf Acknowledgment.} This work was supported in part by the
Sonderforschungsbereich 341 K\"oln-Aachen-J\"ulich. I thank
M.~B\"uttiker and H.A.~Weidenm\"uller for useful discussions.
\vfill\eject

\noindent{\bf Appendix A}\medskip

To establish Eq.~(17) on a rigorous level, we find it convenient to
make use of the elegant and powerful technique of Efetov [21], who
has shown that $P(x,y) = \langle |G^+(x,y)|^2 \rangle$ can -
in the diffusive regime - be regarded as the propagator
of certain $Q$-matrix superfields with ``free energy''
functional ${\cal F}[Q]$. It is not difficult to adapt Efetov's
treatment to the reduced model with boundary conditions (12) or
the alternative formulation in terms of the effective Hamiltonian
(14). For simplicity, we assume that electrons are spinless and
subject to potential scattering only. In this case, one finds that
Eq.~(3.53) of Ref.~[21] is replaced by
	$$
	{\cal F}[{Q}] = - {\hbar\pi\nu\over 8\tau}
	\int_{V_0} {\rm str} {Q}^2(x) d^3x
	+ {1\over 2} \int_{V_0} \left( {\rm str}\ln
	\bigl( E_F - {p^2\over 2m} + {i\hbar\over 2\tau}
	{Q} + {i\hbar\over 2}\beta\Lambda\delta_C
	\bigr) \right) (x,x) d^3x ,
	$$
and $P(x,y)$ is a sum of diffuson ($d$) and cooperon
($c$) contributions, $P = P_d + P_c$, where
	$$\eqalignno{
	P_d(x,y) &= \int {\cal D}{Q} ~ {\cal G}_{d}^{+-}
	(x,x;{Q}) {\cal G}_{d}^{-+}(y,y;{Q})
	e^{-{\cal F}[{Q}]},			&{\rm (A.1a)}\cr
	P_c(x,y) &= \int {\cal D}{Q} ~ {\cal G}_{c}^{+-}
	(x,y;{Q}) {\cal G}_{c}^{-+}(y,x;{Q})
	e^{-{\cal F}[{Q}]},			&{\rm (A.1b)}\cr}
	$$
and
	$$
	{\cal G}(x,y;{Q}) = \left( E_F - {p^2\over 2m}
	+{i\hbar\over 2\tau}{Q} + {i\hbar\over 2}\beta
	\Lambda\delta_C \right)^{-1} (x,y).
	$$
The notations used are either self-explanatory or those of
Ref.~[21]. In the next step,
one makes the usual saddle-point approximation followed by a
gradient expansion [21], valid in the metallic regime
$k_F \ell \gg 1$. Retaining in ${\cal F}$ only terms up to
linear order in  $\beta$, one obtains
	$$
	{\cal F}[Q] \simeq {\hbar\pi\nu\over 8} \left(
	D \int_{V_0} {\rm str}(\nabla Q)^2 d^3x + 2
	\int_C \beta{\rm str}\Lambda Q d^2x \right)
	\eqno({\rm A.2})
	$$
where $Q$ satisfies the nonlinear constraint $Q^2 = 1$.
Eq.~(A.2) omits such terms as
	$$
	{\rm const} \times (\hbar\pi\nu)^2 \ell^2 \int_C
	\beta^2 {\rm str}(\Lambda Q)^2 d^2x .
	$$
These are negligible under the locally-weak absorption condition (16).
With the same degree of accuracy, one may use the approximation
${\cal G}(x,x;{Q}) \simeq -i\pi\nu Q(x)$, which leads to
	$$
	P_d(x,y) = - \pi^2 \nu^2 \int {\cal D}Q ~
	Q_{d}^{+-}(x) Q_{d}^{-+}(y) e^{-{\cal F}[Q]}.
	$$
To analyze this expression further, one observes that the
condition $\tau_{\rm diff} \ll \tau_{\rm decay}$ for a
conductor to be decay-width dominated can be written
	$$
	{hD\over L^2} \gg {\hbar \over \rm vol}
	\int_C \beta d^2x .
	$$
A glance at Eq.~(A.2) then shows that the dominant contribution
to $P_d(x,y)$ comes from {\it spatially constant} $Q$-fields,
spatially varying $Q$-fields being separated from the constant
ones by a large gap in free energy. This proves the independence
of $P_d(x,y)$ of $x$ and $y$. \medskip

The other properties used in Sect.~5 are $P_c(x,x) \sim P_d$
and the short range of $P_c(x,y)$. These follow from Eq.~(A.1b) and
	$$
	{\cal G}(x,y;Q) \simeq \left( E_F - {p^2\over 2m} +
	{i\hbar\over 2\tau} Q \right)^{-1} (x,y)
	\simeq f(x-y) Q\bigl((x+y)/2\bigr)
	$$
where $f(0) = -i\pi\nu$, and
$f$ decays over a length scale of the order of the elastic
mean-free path. \medskip

Finally, we observe that the various perturbations in $H$, Eq.~(2),
which break time-reversal and spin-rotation symmetry,
do not affect the diffuson
degrees of freedom, whose free energy is always given by a functional
of the form (A.2). Therefore, $P_d$ is constant in general. \bigskip

\noindent{\bf Appendix B}\medskip

With $\Gamma = \hbar \int_C \beta d^2x / {\rm vol}$ the
decay width, let $\rho_\Gamma(E)$,
	$$
	\rho_\Gamma(E) = - {1\over \pi} {\rm Im}~{\rm tr}
	(E+i\Gamma/2-H)^{-1},
	$$
be the total density of states (counting spin) of the open system.
We will write down exact expressions for
the two-level correlation function
	$$
	R(E_1,E_2;\Gamma) = {\langle \rho_\Gamma(E_1) \rho_\Gamma(E_2)
	\rangle \over \langle\rho_\Gamma(E_1)\rangle\langle
	\rho_\Gamma(E_2)\rangle} - 1
	$$
for all universality classes, obtained by analytic continuation of
Efetov's results [22]. \medskip

We introduce the functions $({\bf C}\to{\bf C})$
	$$\eqalignno{
	f(z) &= - {e^{i\pi z}\sin\pi z \over i\pi^2 z^2},&{\rm (B.1a)}\cr
	g(z) &= \left( {1 \over \pi} {\partial\over\partial z}
	{e^{i\pi z}\over i\pi z}\right) \int_0^1
	{\sin\pi zt \over t} {dt},			&{\rm (B.1b)}	\cr
	h(z) &= \left( - {1\over \pi}{\partial\over\partial z}
	{\sin\pi z\over\pi z}\right) \int_1^\infty
	{e^{i\pi zt} \over it} {dt}			&{\rm (B.1c)}	\cr}
	$$
and, with $\Delta^{-1} = \langle \rho_\Gamma(E_F) \rangle$, we set
	$$\eqalignno{
	&F_{\rm I}(z\Delta) = f(z/2) + g(z/2) ,			\cr
	&F_{\rm IIa}(z\Delta) = f(z/2),
	\quad F_{\rm IIb}(z\Delta) = f(z),	&({\rm B}.2)	\cr
	&F_{\rm III}(z\Delta) = f(z) + h(z) .			\cr}
	$$
$R$ is then given by
	$$
	R_k(E_1,E_2;\Gamma) = {1\over 2} \bigl(
	F_k(E_1-E_2+i\Gamma) + F_k(E_2-E_1+i\Gamma) \bigr)
	\eqno({\rm B}.3)
	$$
in all cases ($k$ = I, IIa, IIb, III). \medskip

Note that the function $f$, Eq.~(B.1a), is holomorphic for $z \in {\bf
C}-\{0\}$ and vanishes as $z^{-2}$ at infinity in the upper half of the
complex plane. Hence, by Cauchy's theorem,
	$$
	\int_{\bf R} f(x+i\gamma) dx = 0 \quad {\rm for}~~ \gamma > 0.
	$$
The same statements apply to the functions $g$ and $h$, Eqs.~(B.2b) and
(B.2c), which leads to the sum rule
	$$
	\int_{\bf R} R(E,E';\Gamma) dE' = 0.
	$$
The result (24) for $R(E_1,E_2;\Gamma)$ quoted in Sect.~6
follows from the asymptotic expansions
	$$\eqalignno{
	f(i\gamma) &= {1 \over 2\pi^2 \gamma^2}
	\Bigl(1-e^{-2\pi\gamma}\Bigr),		&({\rm B.4a})\cr
	g(i\gamma) &= {1 \over 2\pi^2 \gamma^2} \left(
	1 + {2\over\pi\gamma} + O(\gamma^{-2}) \right),	&({\rm B.4b})\cr
	h(i\gamma) &= {1 \over 2\pi^2 \gamma^2} \left(
	1 - {2\over\pi\gamma} + O(\gamma^{-2}) \right),	&({\rm B.4c})\cr}
	$$
for $\gamma \gg 1$. Note that Eq.~(B.4a) shows the corrections to
$R_k$, Eq.~(24), for $k = {\rm IIa}$ and IIb to be exponentially small and
therefore not calculable by perturbation expansion in
$(E_1-E_2+i\Gamma)^{-1}$. Furthermore, from the exact expressions
(B.1-3) one can obtain the conductance fluctuations $\langle \delta g_{ab}
\delta g_{cd} \rangle$ for small values of $\Gamma$, which are not
accessible by perturbation theory either. Both $f(i\gamma)$ and
$g(i\gamma)$ behave for $\gamma\to 0$ as $(\pi\gamma)^{-1}$, while
$h(i\gamma) \simeq {1\over 3}\pi\gamma \ln \pi\gamma$. Insertion
of these limiting forms into Eqs.~(B.2), (B.3) and (22a) gives
	$$
	\langle \delta g_{ab} \delta g_{cd} \rangle \simeq
	{\rm const} \times p_a p_b p_c p_d \times \Gamma/\Delta .
	$$
This shows that the conductance fluctuations go to zero for $\Gamma\to
0$, which is the limit of an isolated system. In other words, the behavior of
$\langle \delta g_{ab} \delta g_{cd} \rangle$ is dominated by the
decrease of $\langle g_{ab} \rangle \langle g_{cd} \rangle \sim
\Gamma^2/\Delta^2$ for $\Gamma\to 0$, overpowering the increase in
{\it relative}
fluctuations due to the appearance of isolated resonances.
This conclusion is contrary to the one reached in Ref.~[4].
We attribute the discrepancy to overextension, in Ref.~[4], of
perturbation theory to a regime where it does not apply. \vfill\eject

\noindent{\bf References}\medskip

\noindent
\item{[1]}
S.~Washburn and R.A.~Webb, Adv.~Phys.~{\bf 35} (1986) 375.

\noindent
\item{[2]}
P.A.~Lee and A.D.~Stone, Phys.~Rev.~Lett.~{\bf 55} (1985) 1622;
B.L.~Al'tshuler, Pis'ma Zh.~Eksp.~Teor.~Fiz.~{\bf 41} (1985) 530
[JETP Lett.~{\bf 41} (1985) 648].

\smallskip\noindent
\item{[3]}
P.A.~Lee, A.D.~Stone and H.~Fukuyama, Phys.~Rev.~B{\bf 35}
(1987) 1039.

\smallskip\noindent
\item{[4]}
R.A.~Serota, S.~Feng, C.~Kane and P.A.~Lee, Phys.~Rev.~B{\bf 36}
(1987) 5031.

\smallskip\noindent
\item{[5]}
S.~Iida, H.A.~Weidenm\"uller and J.A.~Zuk,
Ann.~Phys.~{\bf 200} (1990) 219;
A.~Altland, Z.~Phys.~B{\bf 82} (1991) 105.

\smallskip\noindent
\item{[6]}
H.A.~Weidenm\"uller, Physica A{\bf 167} (1990) 28.

\smallskip\noindent
\item{[7]}
P.L.~Kapur and R.E.~Peierls, Proc.~Roy.~Soc. (London) {\bf A166} (1938) 277.

\smallskip\noindent
\item{[8]}
L.~Eisenbud and E.P.~Wigner, {\it Nuclear Structure}, Princeton
University Press, Princeton (1958).

\smallskip\noindent
\item{[9]}
A.M.~Lane and R.G.~Thomas, Rev.~Mod.~Phys.~{\bf 30} (1958) 257.

\smallskip\noindent
\item{[10]}
C.~Mahaux and H.A.~Weidenm\"uller, {\it Shell-Model Approach to
Nuclear Reactions}, North-Holland Publ. Co., Amsterdam (1969).

\smallskip\noindent
\item{[11]}
H.L.~Harney and A.~H\"upper, Z.~Phys.~{\bf A328} (1987) 327;
H.L.~Harney, A.~H\"upper, M.~Mayer and A.~M\"uller, Z.~Phys.~{\bf
A335} (1990) 293; A.~M\"uller and H.L.~Harney,
Z.~Phys.~{\bf A337} (1990) 465.

\smallskip\noindent
\item{[12]}
P.A.~Lee and T.V.~Ramakrishnan, Rev.~Mod.~Phys.~{\bf 57} (1985) 287.

\smallskip\noindent
\item{[13]}
H.~Baranger and A.D.~Stone, Phys.~Rev.~B{\bf 40} (1989) 8169.

\smallskip\noindent
\item{[14]}
M.~Janssen, Sol. State Commun. {\bf 79} (1991) 1073.

\smallskip\noindent
\item{[15]}
M.C.~Gutzwiller, {\it Chaos in Classical and Quantum Mechanics},
Springer, New York (1990).

\smallskip\noindent
\item{[16]}
R.P.~Feynman and A.R.~Hibbs, {\it Quantum Mechanics and Path
Integrals}, McGraw-Hill, New York (1965).

\smallskip\noindent
\item{[17]}
G.~Bergmann, Phys.~Rep.~{\bf 107} (1984) 1.

\smallskip\noindent
\item{[18]}
W.~Hauser and H.~Feshbach, Phys.~Rev.~{\bf 87} (1952) 366.

\smallskip\noindent
\item{[19]}
B.L.~Al'tshuler and B.I.~Shklovskii, Zh.~Eksp.~Teor.~Fiz.~{\bf 91}
(1986) 220 [Sov.~Phys. JETP~{\bf 64} (1986) 127].

\smallskip\noindent
\item{[20]}
S.~Hikami, A.I.~Larkin and Y.~Nagaoka, Prog.~Theor.~Phys.~{\bf 63}
(1980) 707.
\smallskip\noindent
\item{[21]}
K.B.~Efetov, Adv.~Phys.~{\bf 32} (1983) 53.

\smallskip\noindent
\item{[22]}
K.B.~Efetov, Sov.~Phys.~JETP~{\bf 56} (1982) 467.

\smallskip\noindent
\item{[23]}
J.J.M~Verbaarschot, H.A.~Weidenm\"uller, M.R.~Zirnbauer,
Phys.~Rep. {\bf 129} (1985) 367.
\vfill\eject
\end